\documentclass[14pt]{article}
\usepackage{adjustbox} 
\usepackage[margin=1in]{geometry}
\usepackage{setspace}            
\usepackage{orcidlink}
\usepackage{array}
\usepackage{enumitem}
\usepackage{lineno}
\usepackage{pbox}
\usepackage{amssymb,amsmath,amsthm,latexsym}
\usepackage{color}
\usepackage{xcolor}
\usepackage{tablefootnote}
\usepackage{booktabs}
\usepackage{siunitx,caption}
\usepackage{footnote}
\usepackage{graphicx}
\usepackage{float}
\usepackage[labelformat=simple]{subfig}

\usepackage{lipsum}
\usepackage[para]{threeparttable}
\usepackage{tikz}
\usepackage{endnotes}
\usepackage[toc]{appendix}
\usepackage{caption}
\usepackage{lscape}
\usepackage{fancybox,fancyvrb}
\usepackage{verbatimbox}
\usepackage{pxfonts}
\usepackage{url}

\newtheorem{thm}{Theorem}[section]

\newtheorem{rema}[thm]{Remark}
\newtheorem{defi}[thm]{Definition}

{\theoremstyle{remark}}

\usetikzlibrary{arrows}

\title{\textbf{CMHSU: An R Statistical Software Package to Detect Mental Health Status,
Substance Use Status, and their Concurrent Status
in the North American Healthcare Administrative Databases}}

\author{
  Mohsen Soltanifar, Ph.D, P.Stat.\orcidlink{0000-0002-5989-0082}\thanks{
    $^{1}$DARE Department, BC PHSA Corporate, 1333 West Broadway, Vancouver, BC V6H 1G9, Canada; \\
    $^{3}$College of Professional Studies, Northeastern University, 410 W Georgia St, Vancouver, BC V6B1Z3, Canada.
  }
  \quad and \quad
  Chel Hee Lee, Ph.D\orcidlink{0000-0001-8209-8176}\thanks{
    $^{2}$CCM Department, AHS Corporate, 3260 Hospital Drive NW, Calgary, AB T2N 4Z6, Canada;\\
    $^{4}$Mathematics\&Statistics Department, University of Calgary, 2500 University Drive NW, Calgary, AB T2N 1N4, Canada.
    \newline
    \textbf{Corresponding author:} \texttt{m.soltanifar@northeastern.edu}
  }
    }

\date{\vspace{0.5em} April 06, 2025}

\begin{document}
\maketitle

\begin{abstract}
\noindent
The concept of concurrent mental health and substance use (MHSU) and its detection in patients has garnered growing interest among psychiatrists and healthcare policymakers over the past four decades. Researchers have proposed various diagnostic methods, including the Data-Driven Diagnostic Method (DDDM), for the identification of MHSU. However, the absence of a standalone statistical software package to facilitate DDDM for large healthcare administrative databases has remained a significant gap. This paper introduces the R statistical software package \texttt{CMHSU}, available on the Comprehensive R Archive Network (CRAN), for the diagnosis of mental health (MH), substance use (SU), and their concurrent status (MHSU). The package implements DDDM using hospital and medical service physician visit counts along with maximum time span parameters for MH, SU, and MHSU diagnoses.  A simulated real-world dataset incorporating fentanyl is presented to examine various analytical aspects, including three key dimensions of MHSU detection based on the DDDM framework, as well as temporal analysis to demonstrate the package's application for healthcare policymakers. Additionally, the limitations of the \texttt{CMHSU} package and potential directions for its future extension are discussed.
\end{abstract}

\vspace{1em}

\vspace{1em}
\begin{quote}
\emph{Integrating mental health and substance use data within our healthcare systems is essential for understanding the complex interplay between these disorders and for developing comprehensive treatment strategies that address the full spectrum of patient needs.}

\smallskip
Nora D. Volkow, MD (2021)\\
National Institute on Drug Abuse of the United States
\end{quote}

\section{Introduction}\label{sec1}
The introduction of this paper is organized into four main sections: Section \ref{sec1.1} outlines definitions and diagnostic approaches for Concurrent Mental Health and Substance Use status (MHSU), with a particular emphasis on the Data-Driven Diagnostic Method (DDDM). Section \ref{sec1.2} provides a summary of the primary North American healthcare administrative databases. Section \ref{sec1.3} explores the motivations for developing the ``CMHSU'' statistical software package. Lastly, section \ref{sec1.4} details the outline of the study.

\subsection{Concurrent Mental Health Status and Substance Use Status: Definition \& Diagnosis}\label{sec1.1} 

The concept of concurrent mental health (MH) and substance use (SU) status, collectively referred to as MHSU, has been explored for over four decades \cite{Weiss1985,Jaffe1986,Helzer1988,Drake1989,Regier1990,Drake1993,Kessler1994,Wright2000,Evans2001,Cantwell2003,Blanco2012,Szerman2012,Atkins2014}, with significant attention given to its methodological challenges \cite{Todd2004}. Various approaches have been developed to diagnose MHSU, each with distinct applications and strengths. The primary diagnostic methods are outlined below:

\subsubsection{Comprehensive Clinical Assessment (CCA)}
This method involves in-depth evaluations and interviews conducted by healthcare professionals to gain a thorough understanding of an individual’s mental health, substance use patterns, and psychosocial factors. CCA typically includes assessments of psychiatric and medical history, as well as risk evaluation \cite{Mueser2003,Sciacca1996}. For example, a psychiatrist may utilize the Psychiatric Research Interview for Substance and Mental Disorders (PRISM) to diagnose major depressive disorder and evaluate alcohol dependency patterns.

\subsubsection{Standardized Screening Instruments (SSI)}
SSIs are structured tools designed to assess the presence and severity of mental health and substance use disorders. Common instruments include the Generalized Anxiety Disorder-7 (GAD-7), the Patient Health Questionnaire-9 (PHQ-9), the Drug Abuse Screening Test (DAST), and the Alcohol Use Disorders Identification Test (AUDIT) \cite{Skinner1982,Babor2001}. For instance, a general practitioner might employ the AUDIT to detect harmful drinking behaviors in a patient suspected of having an anxiety disorder.

\subsubsection{Integrated Treatment Approach (ITA)}
ITA emphasizes the simultaneous treatment of mental health and substance use disorders within a unified, evidence-based framework. This approach prioritizes coordinated strategies to address the complex interactions between these conditions \cite{Drake2008,Minkoff2001}. For example, a patient with bipolar disorder and opioid use disorder may receive medication-assisted treatment (MAT) alongside cognitive-behavioral therapy (CBT) as part of an integrated care plan.

\subsubsection{Multidisciplinary Collaboration (MC)}
This method involves a coordinated effort among a team of professionals, such as psychiatrists, addiction specialists, and social workers, to design and implement a comprehensive care plan tailored to individuals with co-occurring disorders \cite{Sterling2011,Mueser2002}. For instance, a multidisciplinary team comprising a psychiatrist, a social worker, and an addiction counselor may convene weekly to plan and manage the care of a patient diagnosed with schizophrenia and alcohol dependence.

\subsubsection{Data-Driven Diagnostic Method (DDDM)}
The DDDM leverages clinical and administrative data analysis to identify co-occurring mental health and substance use status through the utilization of healthcare databases and the International Classification of Disease (ICD) codes. This approach is based on patterns of healthcare usage, such as hospitalizations and physician visits \cite{Heslin2015,Keen2022,Lavergne2022a,Lavergne2022b}. For example, a healthcare researcher may analyze hospital records to identify individuals who, within the past year, had at least one hospitalization or two medical physician visits for mental health and substance use status, as indicated by ICD-10 codes \cite{Keen2022,Lavergne2022a,Lavergne2022b}. 

\begin{figure}[H]
\centering
\includegraphics[clip,width=1.0\columnwidth]{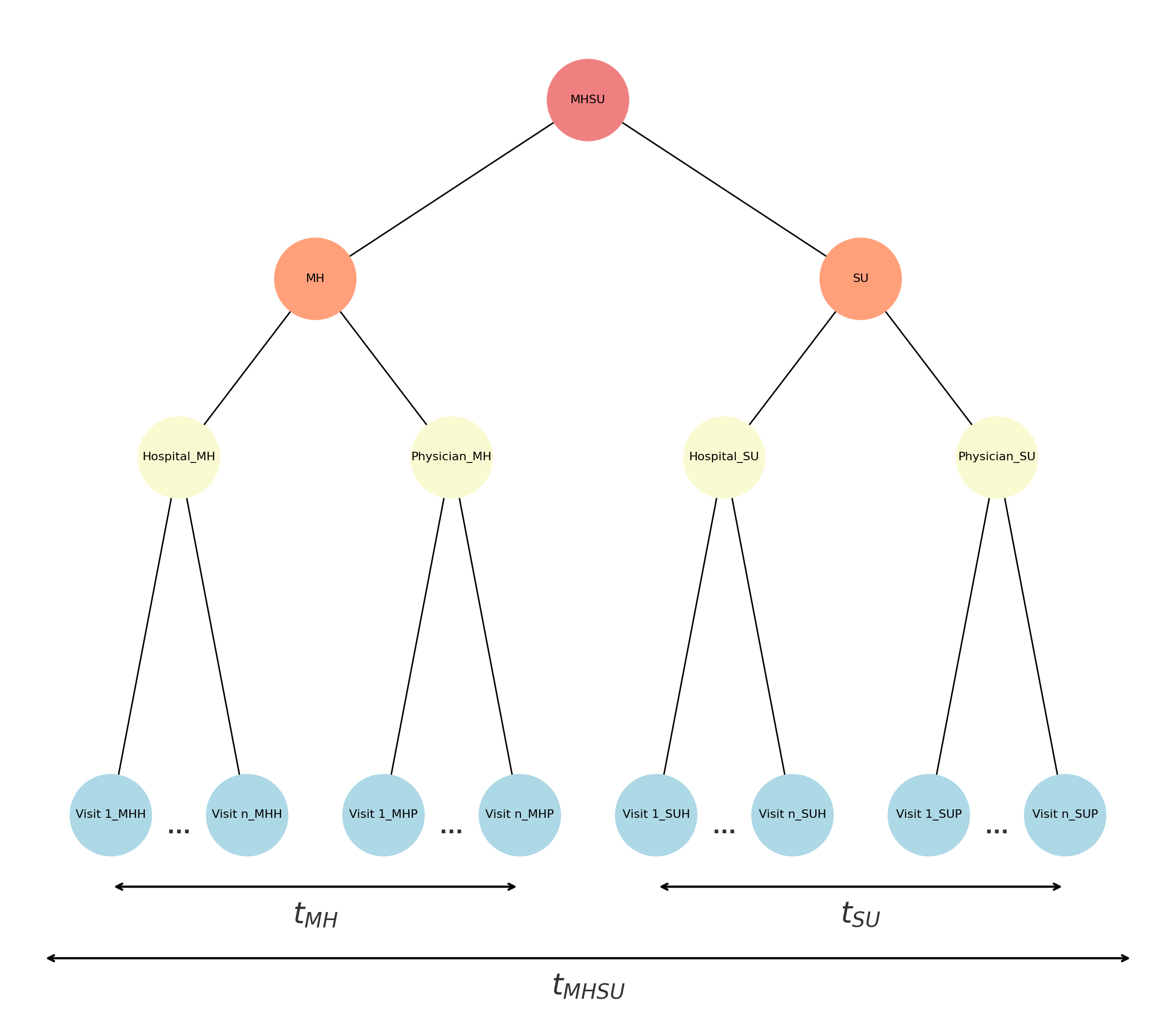}
\caption{Diagram of detection of Mental Health (MH) status, Substance Use (SU) status and their concurrent (MHSU) status in the North American healthcare administrative databases in terms of a given number of hospital visits, medical service physician visits, and time intervals within and between them.\label{fig1}}
\end{figure}

\smallskip
The DDDM approaches have been based on frequency and time-based detection of plausible ICD codes from healthcare administrative databases \cite{Keen2022,Lavergne2022a,Lavergne2022b}. In general format, we set the following general parametric definition \cite{Khan2017,HealthCanada2022,ALHS2011} (Figure \ref{fig1}):

\begin{defi}\label{def1}
\textbf{Data-Driven Diagnostic Method (DDDM)} \\
Given time spans  $t_{MH},t_{SU},$ and $t_{MHSU}$ and frequencies $n_{MHH},n_{MHP},n_{SUH},n_{SUP}.$ Then, a patient has concurrent Mental Health Substance Use (MHSU) diagnosis if and only if in the past $t_{MHSU}$ time span the patient had the following two conditions:
\begin{enumerate}[label={[\alph*]}]
\item at least one Mental Health (MH) diagnosis (defined by at least $n_{MHH}$ times hospitalizations or $n_{MHP}$ primary care physician visits within $t_{MH}$ time span),
\item at least one Substance Use (SU) diagnosis (defined by at least $n_{SUH}$ times hospitalizations or $n_{SUP}$ primary care physician visits within $t_{SU}$ time span). 
\end{enumerate}
\end{defi}

\begin{rema}
The diagnosis of MH status or SU status is based on detection of associated ICD-09, ICD-10, or ICD-11 codes from administrative databases. Examples are provided in Section \ref{sec3} with a simulated database. 
\end{rema}

\begin{rema}
The precise values for the set of parameters $t_{MH},t_{SU}, t_{MHSU},  n_{MHH},n_{MHP},n_{SUH},n_{SUP}$ is decided by the study’s principal clinician, given various internal, external, and contextual factors.
\end{rema}

\subsection{North American Healthcare Administrative Databases}\label{sec1.2} 

In both Canada and the United States, numerous health care administrative databases are available, varying by regional level (federal/state or provincial) and by the type of institution managing the data (hospitals or medical service physician providers). Table \ref{tab1} outlines the most significant databases across North America. These databases commonly record several crucial variables, including \textit{Client ID}, \textit{Visit Date}, \textit{Discharge Date}, and \textit{Disease Diagnostic Code}. 

\begin{table}[htbp]
\centering
\caption{The North American healthcare administrative databases in terms of country, level, hospital, and medical service physician categories.\label{tab1}}
\begin{adjustbox}{max width=\textwidth, max totalheight=0.8\textheight, keepaspectratio}
\begin{tabular}{p{3.5cm}p{1.5 cm}p{8.5cm} p{8.5cm}} 
\toprule
\textbf{Country} & \textbf{Level} & \textbf{Hospital Data} & \textbf{Physician Data}\\
\midrule
\textbf{Canada} (\raisebox{-0.25\height}{\includegraphics[height=12pt]{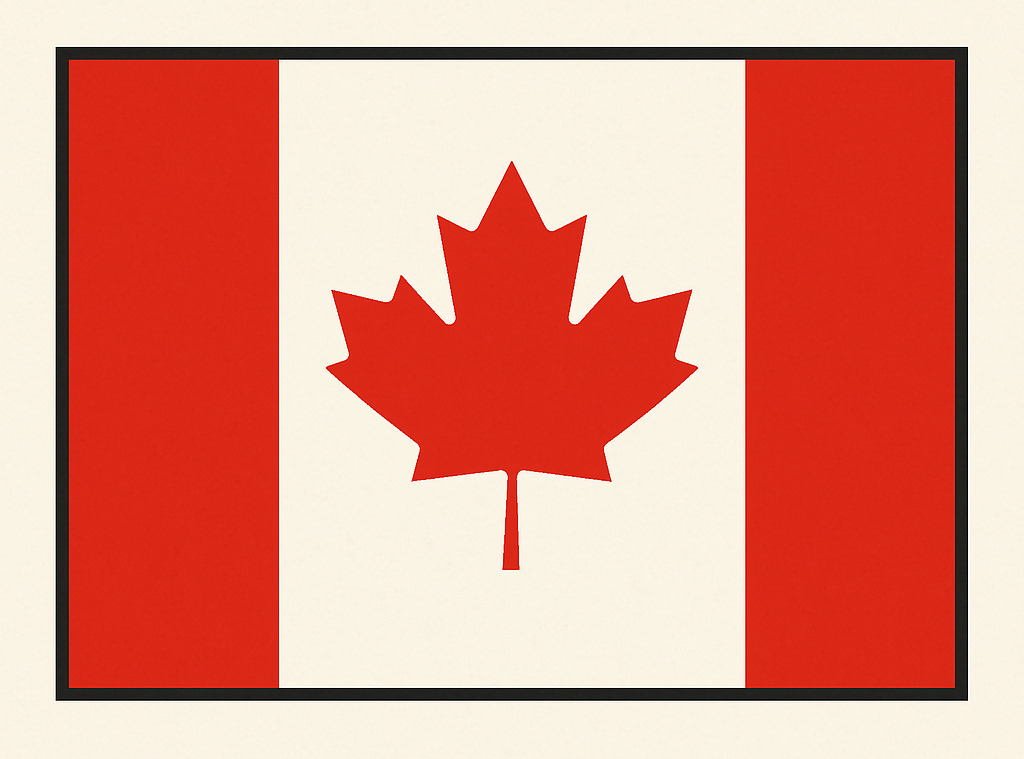}})
 & \textbf{Federal} &
\begin{itemize}[leftmargin=*]
    \item Discharge Abstract Database (DAD)
    \item National Ambulatory Care Reporting System (NACRS) 
\end{itemize}
&
\begin{itemize}[leftmargin=*]
    \item Canadian Institute of Health Information(CIHI) Physician Claims Data
    \item Canadian Management Information Database (CMDB)
\end{itemize}
\\
\cmidrule(lr){2-4}
 & \textbf{Province} &
\begin{itemize}[leftmargin=*]
    \item MED-ÉCHO (Quebec) data
    \item Ontario Health data
    \item Alberta Health data
\end{itemize}
&
\begin{itemize}[leftmargin=*]
    \item Medical Services Plan (MSP) Data (BC)
    \item Alberta Health Physician Claims
    \item Régie de l'assurance maladie du Québec (RAMQ)
    \item Ontario Health Insurance Plan (OHIP)
\end{itemize}
\\
\cmidrule(lr){2-4}
 & \textbf{Territory} &
\begin{itemize}[leftmargin=*]
    \item Yukon Hospital Corporation Dis-charge Data
    \item Northwest Territories Health Authority Records
\end{itemize}
&
\begin{itemize}[leftmargin=*]
    \item Yukon Health Care Insurance Plan Claims
    \item Northwest Territories Medical Billing Data
\end{itemize}
\\
\midrule
\textbf{United States} (\raisebox{-0.25\height}{\includegraphics[height=12pt]{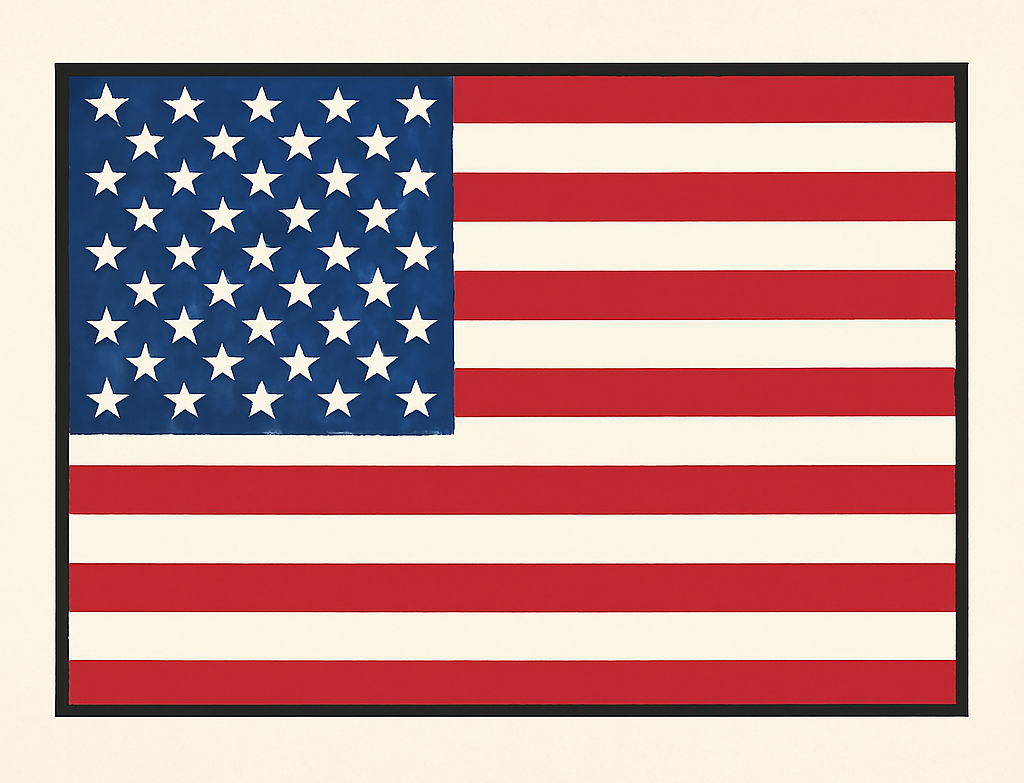}})
 & \textbf{Federal} &
\begin{itemize}[leftmargin=*]
    \item National Inpatient Sample (NIS)
    \item Medicare Provider Analysis and Review (MEDPAR)
    \item Veterans Affairs (VA) Inpatient Data
\end{itemize}
&
\begin{itemize}[leftmargin=*]
    \item Medicare Physician/Supplier Part B Claims Data
    \item National Ambulatory Medical Care Survey (NAMCS)
    \item Veterans Affairs (VA) Outpatient Data
\end{itemize}
\\
\cmidrule(lr){2-4}
 & \textbf{State} &
\begin{itemize}[leftmargin=*]
    \item State Inpatient Databases (SID)
    \item California Office of Statewide Health Planning
and Development (OSHPD)
    \item New York Statewide Planning and Research
Cooperative System (SPARCS)
\end{itemize}
&
\begin{itemize}[leftmargin=*]
    \item State Medicaid Claims Data
    \item All-Payer Claims Databases (APCDs: e.g.,
Massachusetts, Colorado)
\end{itemize}
\\
\cmidrule(lr){2-4}
 & \textbf{Territory} &
\begin{itemize}[leftmargin=*]
    \item Guam Memorial Hospital Authority Records
    \item American Samoa Dept.\ of Health Inpatient Data
\end{itemize}
&
\begin{itemize}[leftmargin=*]
    \item Guam Medicaid and CHIP Claims Data
    \item American Samoa Medicaid Claims\textsuperscript{1}
\end{itemize}
\\
\bottomrule
\\
\end{tabular}
\end{adjustbox}
\smallskip
\noindent
{\\
\footnotesize{\textsuperscript{1} As of January 03, 2025.}}
\end{table}

\noindent
The term \textit{Disease Diagnostic Code} refers to codes from the World Health Organization's International Classification of Diseases (ICD) \cite{WHO2019}. The ICD codes utilized in American and Canadian healthcare systems are detailed in \cite{CIHI2022,CDCP2024}. Within each country, state or provincial systems may adopt their versions of these federally recognized ICD codes \cite{Keen2022,Lavergne2022a,Lavergne2022b,Casillas2022}. In this study, we consider ICD codes broadly, irrespective of their version (ICD-09, ICD-10, or ICD-11), focusing on codes ranging from 3 to 5 characters in length, presented without a decimal point.\par 
\begin{rema}
The ICD codes provided in the following examples for mental health (MH) and substance use (SU) statuses are intended as illustrative examples rather than exhaustive lists. The complete set of ICD codes for these conditions may vary based on local and national guidelines, as well as the criteria established by study principal investigators.
\end{rema}

\subsection{Motivation}\label{sec1.3} 
The DDDM has several advantages over other methods for the diagnosis of MHSU as follows:
\begin{enumerate}[label={(\roman*)}]
\item \textit{Scalability and Large-Scale Analysis}: The Data-Driven Diagnostic Method (DDDM) offers the advantage of scalability by utilizing healthcare databases and administrative records to analyze large populations. Unlike approaches that require direct patient interaction, DDDM enables the comprehensive assessment of co-occurring mental health and substance use status across diverse demographic groups and healthcare systems \cite{Heslin2015}.

\item \textit{Objective and Quantitative Approach}: DDDM relies on the objective extraction and analysis of data, such as ICD codes and patterns of healthcare utilization. This minimizes biases often associated with subjective methods, such as clinical interviews, thereby improving the reliability and validity of diagnostic outcomes \cite{Keen2022}.

\item \textit{Cost-Effectiveness}: By leveraging preexisting healthcare data, DDDM eliminates the need for resource-intensive procedures, such as in-person assessments or multidisciplinary team evaluations. This makes it a cost-efficient alternative for healthcare systems facing resource limitations \cite{Lavergne2022a}.

\item \textit{Timeliness and Accessibility}: DDDM enables rapid identification of MHSU patterns through the querying of existing data sources, providing real-time or near-real-time diagnostic capabilities. This is particularly valuable for tracking trends and addressing emerging public health issues \cite{Lavergne2022b}.

\item \textit{Population-Level Insights for Policy and Planning}: Through its capacity to analyze large-scale healthcare data, DDDM facilitates the identification of utilization trends, disparities, and service gaps. This allows policymakers to design targeted interventions and optimize resource allocation to address the needs of individuals with co-occurring status \cite{Keen2022}.
\end{enumerate}

\noindent
However, despite all the above advantages of DDDM over other methods for diagnosis of MHSU, there has been no available statistical software package with the given parameters in Definition \ref{def1} to help researchers implement MHSU-diagnosis detection. The R Statistical software \texttt{CMHSU} enables researchers to detect MHSU with a variety of scenarios in terms of time span, visit frequency, and subtypes of mental health/substance use.

\subsection{Study Outline}\label{sec1.4}
This paper is structured as follows: Initially, we provide essential theories and instructions for the installation of the \texttt{CMHSU} package. We introduce the four primary functions of the package, which are designed to detect mental health (MH), substance use (SU), and their concurrent status (MHSU) in both basic and comprehensive forms. Subsequently, we describe a simulation study using a simulated real-world healthcare administrative dataset to identify MH, SU, and MHSU statuses. This study comprises four sub-studies. In the first three, specific parameters from Definition \ref{def1} are held constant while others vary, enabling an analysis of trends in the detection of MH, SU, and MHSU statuses. The final sub-study focuses on temporal analysis, demonstrating an example of package application for North American healthcare policymakers. We conclude with a discussion on the contributions of the \texttt{CMHSU} package to existing literature, its limitations, and future development plans.\par 

\section{The CMHSU Package}\label{sec2}
\subsection{The Background and Installation}\label{sec2.1}
The \texttt{CMHSU} package utilizes the Data-Driven Diagnostic Method (DDDM) parametric Definition \ref{def1} detailed in Section \ref{sec1.1} and illustrated in Figure \ref{fig1}. It requires four R packages: \texttt{dplyr} \cite{Wickham2023a}, \texttt{tidyr} \cite{Wickham2023b}, \texttt{purrr} \cite{Wickham2023c}, and \texttt{magrittr} \cite{Bache2022}, and was developed using R version 4.2.2 \cite{RCoreTeam2022}. The package (version 0.0.6.9) is accessible on the Comprehensive R Archive Network (CRAN) at 
\begin{center}
\texttt{https://CRAN.R-project.org/package=CMHSU}\, 
\end{center}
(accessed on 10 January 2025). Installation and loading of the package can be performed using R commands:
\begin{verbatim}
Line #1: > install.packages("CMHSU", dependencies=TRUE)
Line #2: > library(CMHSU) 
\end{verbatim}

\noindent
The \texttt{CMHSU} package includes four primary multivariate functions:
\begin{verbatim}
(1) MH_status(),
(2) SU_status(),
(3) MHSU_status_basic(),
(4) MHSU_status_broad().
\end{verbatim}
The first two functions focus specifically on the individual components of Mental Health (MH) or Substance Use (SU) as depicted in Figure \ref{fig1}. The latter two functions encompass the comprehensive framework shown in Figure \ref{fig1}. Each function is described in detail in the subsequent subsections.

\subsection{Detection of Mental Health Status}\label{sec2.2}
The function \texttt{MH\_status()} is designed to determine the mental health status of patients based on a predefined list of plausible mental health diagnostic codes provided by clinicians. This function requires five input parameters. The first parameter, \texttt{inputdata}, represents the patient data and is formatted as a dataframe with four essential columns:
\begin{enumerate}
\item \texttt{ClientID}, which uniquely identifies the patient, 
\item \texttt{VisitDate}, indicating the date of the visit,
\item \texttt{Diagnostic\_H}, representing diagnoses made during hospital visits,
\item \texttt{Diagnostic\_P}, reflecting diagnoses made by medical service physicians.
\end{enumerate}
\noindent
The second, third, and fourth parameters, \texttt{n\_MHH}, \texttt{n\_MHP}, and \texttt{t\_MH}, specify the minimum number of hospital visits required, the minimum number of medical physician visits needed, and the maximum allowable time lag (in days) between hospital or physician visits, respectively. Finally, the fifth parameter, \texttt{ICD\_MH}, contains the list of plausible mental health diagnostic codes determined by the study clinicians. This function returns a dataframe matrix including \texttt{ClientID}, earliest date of mental health, latest date of mental health, and the mental health status (Yes/No).

\smallskip
\noindent
As an initial working example, we utilize a simulated real-world dataset, as described in Section \ref{sec3.1}, to identify all patients with a mental health status corresponding to one of the following status: psychotic, mood, anxiety, or neurocognitive disorders. The detection criteria include at least one hospital visit or at least one medical service physician visit, within a maximum time span of two months (60 days). The corresponding R code for implementing this detection is shown below; it generates the results depicted in Figure~\ref{fig3}a:

\begin{verbatim}
Line #1: > myexample<-SampleRWD[,c(1:4)]
Line #2: > SampleMH_1 = MH_status(myexample, n_MHH=1, n_MHP=1, t_MH=60,
                    ICD_MH=c("F060","F063","F064","F067"))
Line #3: > head(SampleMH_1)
\end{verbatim}

\subsection{Detection of Substance Use Status}\label{sec2.3}
The function \texttt{SU\_status()} is similar to \texttt{MH\_status()}. It is designed to determine the substance use status of patients based on a predefined list of plausible substance use diagnostic codes provided by clinicians. This function also requires five input parameters. The first parameter, \texttt{inputdata}, is a dataframe with columns:
\begin{enumerate}
\item \texttt{ClientID},
\item \texttt{VisitDate},
\item \texttt{Diagnostic\_H},
\item \texttt{Diagnostic\_P}.
\end{enumerate}
The second, third, and fourth parameters, \texttt{n\_SUH}, \texttt{n\_SUP}, and \texttt{t\_SU}, specify the minimum number of hospital visits, the minimum number of physician visits, and the maximum time lag (in days) between visits, respectively. The fifth parameter, \texttt{ICD\_SU}, is the list of plausible substance use diagnostic codes. This function returns a dataframe including \texttt{ClientID}, earliest date of substance use, latest date of substance use, and the substance use status (Yes/No).

\smallskip
\noindent
As the second working example, we identify all patients with a substance use status corresponding to one of the following status: alcohol use, fentanyl use, cannabis use, or cocaine use. The detection criteria require at least one hospital visit or one medical service physician visit, within a maximum allowable time span of two months (60 days). Below is the corresponding R code; it generates the results illustrated in Figure~\ref{fig3}b:

\begin{verbatim}
Line #1: > myexample<-SampleRWD[,c(1:4)]
Line #2: > SampleSU_1 = SU_status(myexample, n_SUH=1, n_SUP=1, t_SU=60,
                      ICD_SU=c("F100","T4041","F120","F140"))
Line #3: > head(SampleSU_1)
\end{verbatim}

\subsection{Detection of Concurrent Mental Health \& Substance Use (MHSU) Status -- Part (I)}\label{sec2.4}
The function \texttt{MHSU\_status\_basic()} is designed to detect concurrent MH and SU status. It requires ten input parameters:
\begin{enumerate}
\item \texttt{inputdata} (dataframe with \texttt{ClientID}, \texttt{VisitDate}, \texttt{Diagnostic\_H}, \texttt{Diagnostic\_P}),
\item \texttt{n\_MHH}, 
\item \texttt{n\_MHP}, 
\item \texttt{n\_SUH}, 
\item \texttt{n\_SUP},
\item \texttt{t\_MH}, 
\item \texttt{t\_SU}, 
\item \texttt{t\_MHSU},
\item \texttt{ICD\_MH},
\item \texttt{ICD\_SU}.
\end{enumerate}
\noindent
Here, \texttt{n\_MHH} and \texttt{n\_MHP} are the minimum numbers of hospital and physician visits for MH, \texttt{n\_SUH} and \texttt{n\_SUP} are the minimum numbers for SU, and \texttt{t\_MH}, \texttt{t\_SU} are maximum lags within MH and SU, respectively. The parameter \texttt{t\_MHSU} is the maximum time span between MH and SU statuses. Both \texttt{ICD\_MH} and \texttt{ICD\_SU} are lists of relevant diagnostic codes. The function returns a dataframe with earliest/latest MH dates, MH status, earliest/latest SU dates, SU status, and the concurrent status (Yes/No). It assumes the input data time span is less than or equal to \texttt{t\_MHSU}.

\smallskip
\noindent
As the third working example, we build on the earlier two to detect the Concurrent Mental Health and Substance Use (MHSU) status for clients diagnosed with psychotic, mood, anxiety, or neurocognitive disorders in combination with alcohol, fentanyl, cannabis, or cocaine consumption-related status. We require a minimum of one hospital visit and one physician visit within two months (60 days) for each of MH and SU; the maximum time span between MH and SU statuses is set to one year (365 days). It is important to note that the total dataset time span is 363 days, which is slightly shorter than the specified maximum span of 365 days. The R code is below; results appear in Figure~\ref{fig3}c:

\begin{verbatim}
Line #1: > myexample<-SampleRWD[,c(1:4)]
Line #2: > SampleMHSU_1 = MHSU_status_basic(myexample, n_MHH=1, n_MHP=1, n_SUH=1,
                    n_SUP=1, t_MH=60, t_SU=60, t_MHSU=365,
                    ICD_MH=c("F060","F063","F064","F067"),
                    ICD_SU=c("F100","T4041","F120","F140"))
Line #3: > head(SampleMHSU_1)
\end{verbatim}

\subsection{Detection of Concurrent Mental Health \& Substance Use (MHSU) Status -- Part (II)}\label{sec2.5}
\begin{figure}[H]
\centering
\includegraphics[clip,width=1.0\columnwidth]{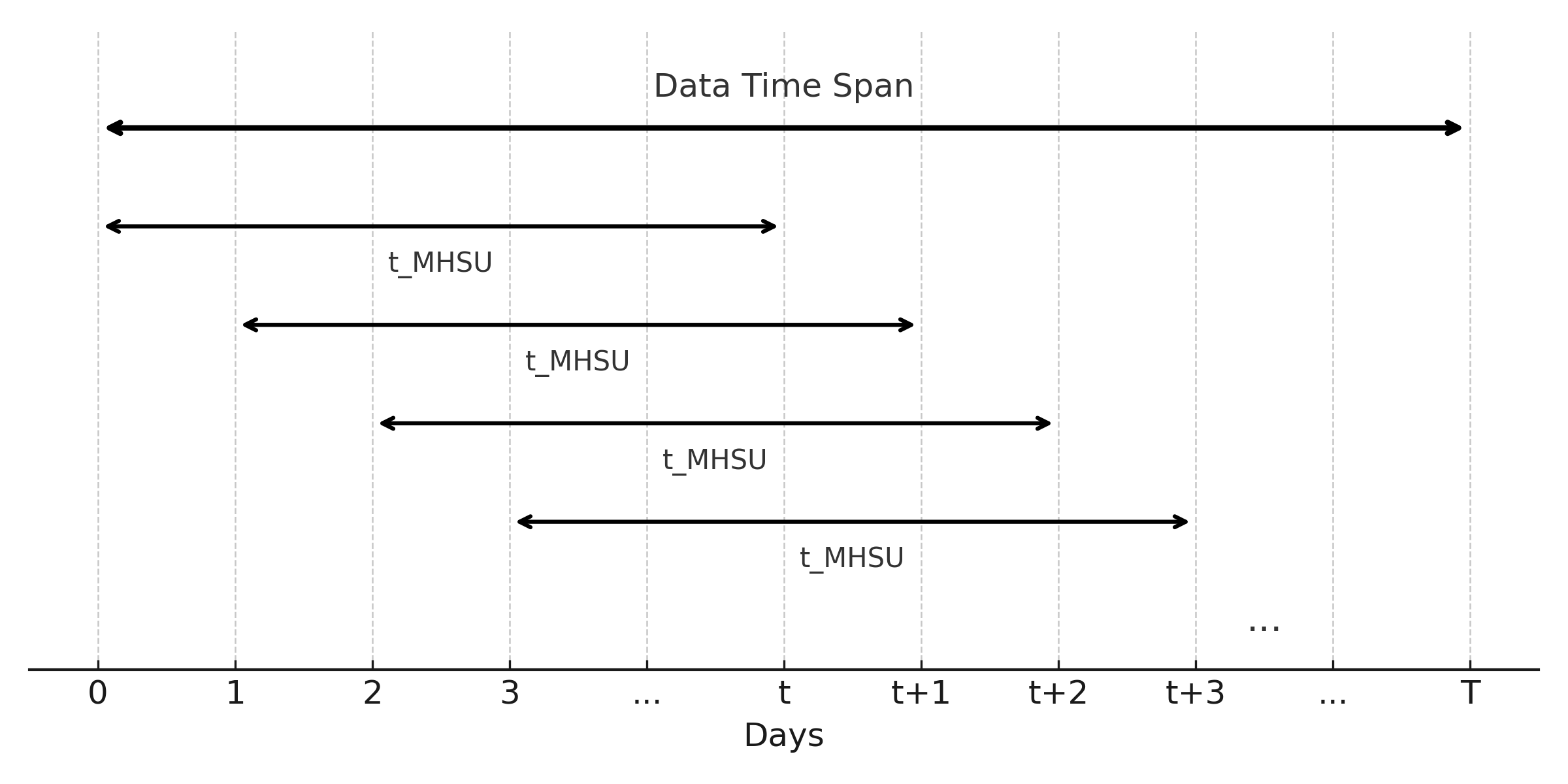}
\caption{The concept of window calculation in function \texttt{MHSU\_status\_broad()}. \label{fig2}}
\end{figure}

A key assumption of \texttt{MHSU\_status\_basic()} is that the data time span is $\leq t\_MHSU$. When this does not hold, \texttt{MHSU\_status\_broad()} can be used. It takes the same ten parameters but generates $k$ non-overlapping windows of concurrent MH and SU output datasets (\,$k = |TimeSpan(\textit{inputdata})| - t_{MHSU} + 1$\,). Each output includes a ``Window'' variable to denote the respective time window (see Figure~\ref{fig2}). Thus, for a dataset of size $n$, the output can have size $k \times n$.

\begin{figure}[H]
\centering
\includegraphics[clip,width=1.0\columnwidth]{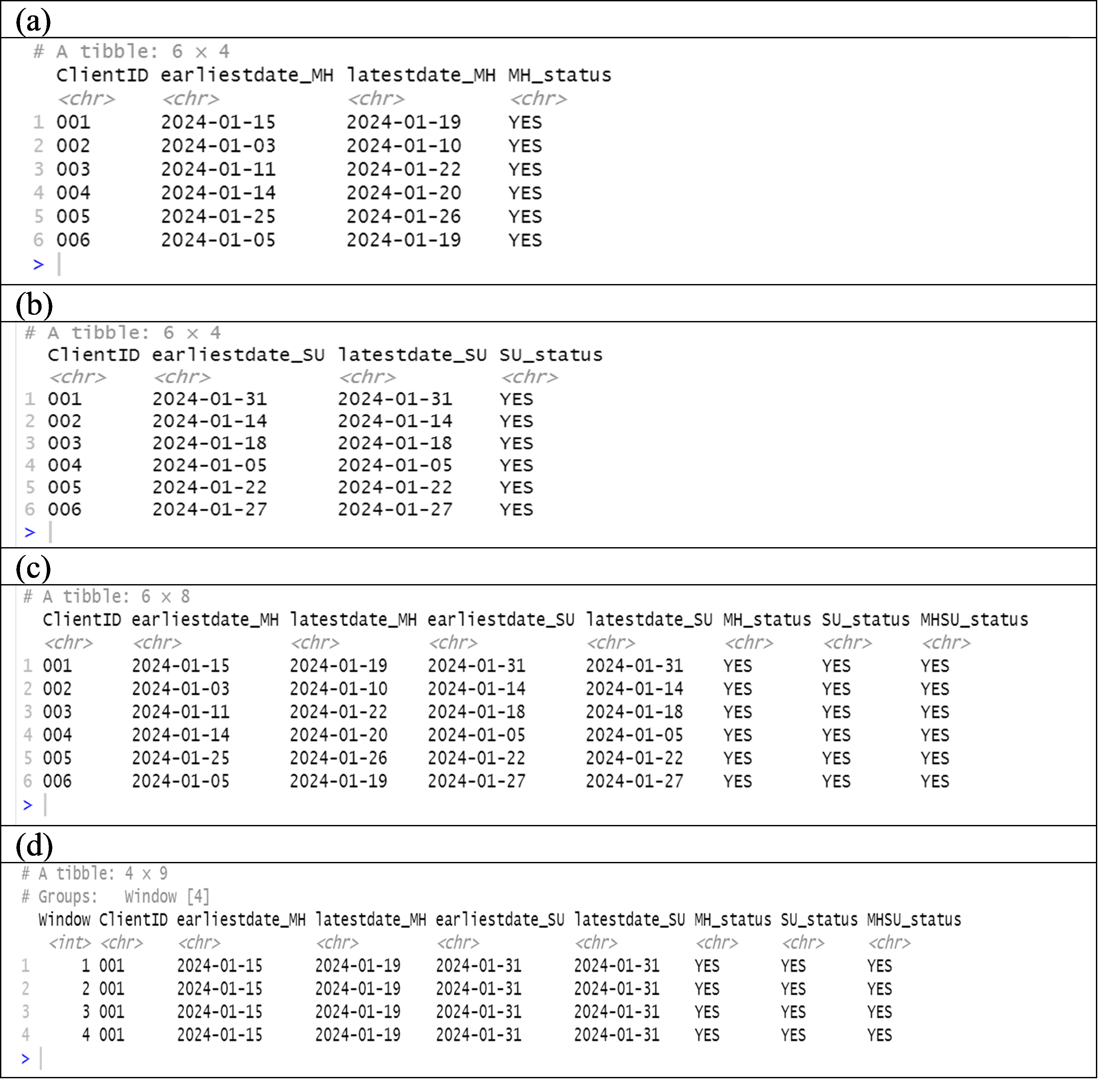}
\caption{Sample Simulated Real World Dataset: (a) Mental Health Status, (b) Substance Use Status, (c) Concurrent Status (basic), and (d) Concurrent Status (broad).\label{fig3}}
\end{figure}

\noindent
As the final working example, we modify the third example to detect MHSU status with a maximum allowable time span between MH and SU (\texttt{t\_MHSU}) of 360 days, while the total dataset is 363 days. This implies $k=363-360+1=4$ non-overlapping windows. The code below illustrates usage; it produces results in Figure~\ref{fig3}d:

\begin{verbatim}
Line #1: > myexample<-SampleRWD[,c(1:4)]
Line #2: > SampleMHSU_2 = MHSU_status_broad(myexample, n_MHH=1, n_MHP=1, n_SUH=1,
                       n_SUP=1, t_MH=60, t_SU=60, t_MHSU=360,
                       ICD_MH=c("F060","F063","F064","F067"),
                       ICD_SU=c("F100","T4041","F120","F140"))
Line #3: > head(SampleMHSU_2[c(1,201,401,601),])
\end{verbatim}

\section{A Simulation Study}\label{sec3}
This section introduces a simulation study, encompassing its data and analysis, and is organized into four key parts. Section \ref{sec3.1} outlines the characteristics of the simulated real-world dataset. Section \ref{sec3.2} examines how varying the maximum time span within MH status and within SU status affects the frequency of MH, SU, and MHSU diagnoses, given the assumption of a constant number of hospital and medical service physician visits. Section \ref{sec3.3} investigates the influence of the number of required hospital and physician visits on MH, SU, and MHSU detection, assuming a fixed maximum time span for MH and SU. Finally, Section \ref{sec3.4} explores the effect of varying the maximum time span within MH, within SU, and between MH and SU on the number of MHSU diagnoses, assuming fixed number of hospital and medical service physician visits. 

\subsection{Simulated Real-World Data}\label{sec3.1}
We simulate a healthcare administrative dataset as follows:
\begin{itemize}
\item The dataset consists of 200 patients categorized into seven diagnostic groups who visited hospitals or medical service physicians from January~1, 2024, to December~31, 2024.
\item The patient groups include: 125 individuals with recorded mental health (MH) diagnoses, 125 with substance use (SU) diagnoses, 100 with concurrent MHSU diagnoses, and 50 with no MH, SU, or MHSU diagnoses.
\end{itemize}

\noindent
Table \ref{tab2} summarizes key features of the simulated dataset, and Table \ref{tab3} provides example entries for two sample patients. This dataset is included as part of the \texttt{CMHSU} R package (see supplementary materials).

\begin{table}[H]
\centering
\caption{Key features of the simulated real-world healthcare administrative database.\label{tab2}}
\begin{adjustbox}{max width=\textwidth, max totalheight=0.8\textheight, keepaspectratio}
\begin{tabular}{ccccccccc}
\toprule
\textbf{Group} & \textbf{Size} & \textbf{Visit Date Span} & \textbf{Visit Length} & \textbf{\#Hospital} & \textbf{\#Physician} & \textbf{SU (freq)} & \textbf{MH (freq)} & \textbf{Other (freq)} \\
\midrule
1 & 10 & 01.01.2024--31.01.2024 & 1 month & 1 & 2 & F100 (1) & F060 (2) & NA \\
2 & 20 & 01.02.2024--31.03.2024 & 2 months & 2 & 4 & T4041 (2) & F063 (4) & J10 (4) \\
3 & 30 & 01.04.2024--31.06.2024 & 3 months & 3 & 6 & F120 (3) & F064 (6) & I10 (3) \\
4 & 40 & 01.07.2024--31.12.2024 & 6 months & 6 & 12 & F140 (6) & F067 (12) & I10 (6), J10 (12) \\
5 & 25 & 01.11.2024--31.12.2024 & 2 months & 3 & 6 & F100 (3) & NA & J10 (6) \\
6 & 25 & 01.11.2024--31.12.2024 & 2 months & 2 & 4 & NA & F060 (4) & I10 (2) \\
7 & 50 & 01.11.2024--31.12.2024 & 2 months & 1 & 2 & NA & NA & I10 (1), J10 (2) \\
\bottomrule
\end{tabular}
\end{adjustbox}
\smallskip
\footnotesize
Notes: F100(Alcohol); F060(Psychotic); T4041(Fentanyl); F063(Mood); F120(Cannabis); F064(Anxiety); F140(Cocaine); F067(Neurocognitive); I10(Hypertension); J10(Influenza); NA(Not Applicable).
\end{table}

\begin{table}[H]
\centering
\caption{Two sample clients from the 200 simulated real-world healthcare administrative database.\label{tab3}}
\begin{adjustbox}{max width=\textwidth, max totalheight=0.8\textheight, keepaspectratio}
\begin{tabular}{cccccccc}
\toprule
\textbf{ClientID} & \textbf{VisitDate} & \textbf{Diagnostic\_H} & \textbf{Diagnostic\_P} & \textbf{MHSU\_H} & \textbf{Meaning\_H} & \textbf{MHSU\_P} & \textbf{Meaning\_P} \\
\midrule
001 & 2024-01-31 & F100 & NA & SU & Alcohol & NA & NA \\
001 & 2024-01-15 & NA & F060 & NA & NA & MH & Psychotic \\
001 & 2024-01-19 & NA & F060 & NA & NA & MH & Psychotic \\
\cmidrule(lr){1-8}
011 & 2024-02-19 & T4041 & NA & SU & Fentanyl & NA & NA \\
011 & 2024-03-07 & T4041 & NA & SU & Fentanyl & NA & NA \\
011 & 2024-02-14 & NA & F063,J10 & NA & NA & MH & Mood,Influenza \\
011 & 2024-02-17 & NA & F063,J10 & NA & NA & MH & Mood,Influenza \\
011 & 2024-03-14 & NA & F063,J10 & NA & NA & MH & Mood,Influenza \\
011 & 2024-03-10 & NA & F063,J10 & NA & NA & MH & Mood,Influenza \\
\bottomrule
\end{tabular}
\end{adjustbox}
\smallskip

\footnotesize
Notes: \texttt{Diagnostic\_H}: ICD code for diagnosis at the Hospital; 
\texttt{Diagnostic\_P}: ICD code for diagnosis by service physician; 
\texttt{MHSU\_H}: MH/SU status assigned at hospital; 
\texttt{MHSU\_P}: MH/SU status assigned by physician; 
\texttt{Meaning\_H}: hospital-assigned ICD meaning; 
\texttt{Meaning\_P}: physician-assigned ICD meaning; 
ICD: International Classification of Diseases.
\end{table}

\subsection{The Impact of Maximum Time Span Within Mental Health and Within Substance Use}\label{sec3.2}
\begin{figure}[H]
\centering
\includegraphics[clip,width=0.85\columnwidth]{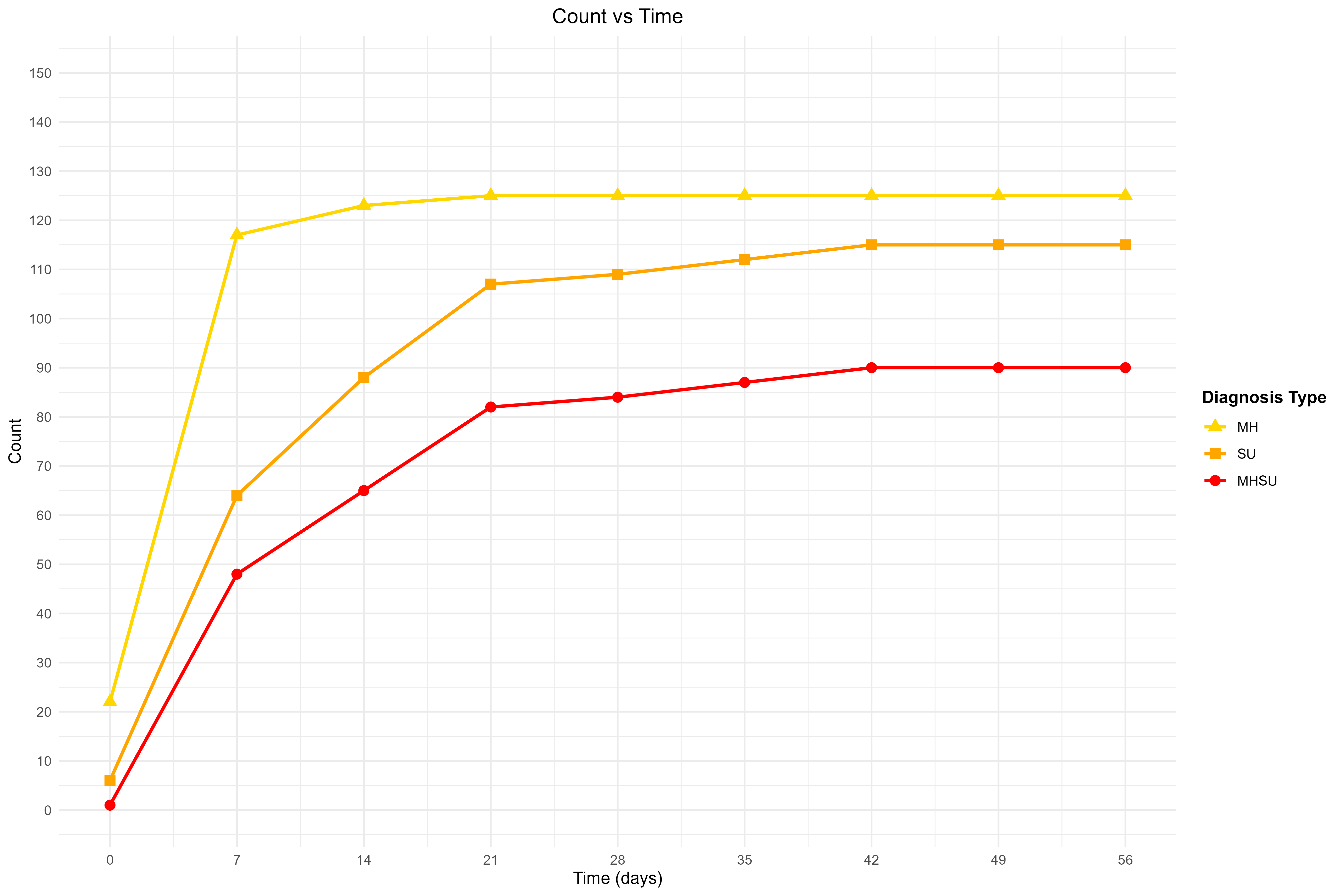}
\caption{Frequency of patients’ diagnosis status vs.\ maximum time span for MH and SU (time = \texttt{t\_MH} = \texttt{t\_SU}, \texttt{t\_MHSU} = 365). \label{fig4}}
\end{figure}

We examine how the maximum allowable time interval within MH and SU diagnoses affects detection counts. Specifically, we require at least two hospital or two physician visits within $x$ days ($x=0\!-\!56$) for both MH and SU, while setting the maximum allowable interval between MH and SU to one year (365 days):
\[
\texttt{n\_MHH} = \texttt{n\_MHP} = \texttt{n\_SUH} = \texttt{n\_SUP}=2,\quad
\texttt{t\_MH}= \texttt{t\_SU}= x \ (0\le x\le 56),\quad
\texttt{t\_MHSU}=365.
\]
Figure~\ref{fig4} presents the results. As the maximum required time interval for diagnosis increases, the number of patients detected rises for MH, SU, and MHSU, eventually reaching an asymptote at 125 patients for MH, 115 patients for SU, and 90 patients for MHSU.

\subsection{The Impact of Number of Hospital Visits and Medical Service Physician Visits}\label{sec3.3}
\begin{figure}[H]
\centering
\includegraphics[clip,width=0.85\columnwidth]{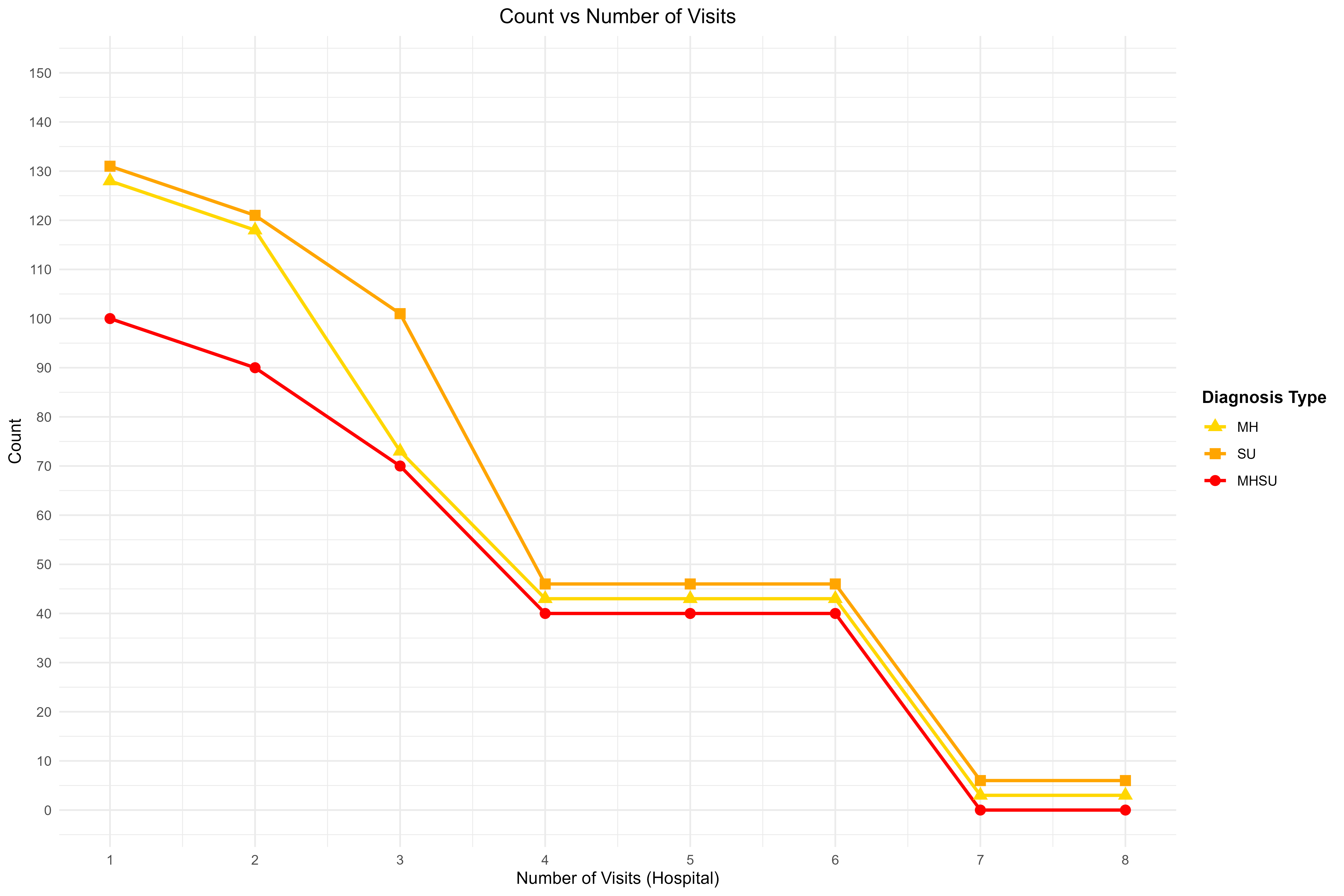}
\caption{Frequency of patients’ diagnosis status vs.\ number of hospital visits, with a hospital:physician count ratio = 1:2 and \texttt{t\_MHSU}=365. \label{fig5}}
\end{figure}

We now investigate how the required number of hospital visits impacts detection counts for MH, SU, and MHSU. We assume a hospital-to-physician visit ratio of 1:2; that is:
\[
\texttt{n\_MHH} = \frac{1}{2}\,\texttt{n\_MHP}, \quad
\texttt{n\_SUH} = \frac{1}{2}\,\texttt{n\_SUP}, \quad
\texttt{n\_MHH} = \texttt{n\_SUH} = x, \quad x=1,\ldots,8,
\]
and fix $\texttt{t\_MH}=\texttt{t\_SU}=183$ (6 months), $\texttt{t\_MHSU}=365$ (1 year). Figure~\ref{fig5} displays the results. As the required number of hospital visits increases, detection counts decrease until eventually reaching zero for all three diagnostic count curves (MH, SU, and MHSU). Additionally, for large required number of hospital visits, the MH and MHSU count curves overlap.

\subsection{The Impact of Maximum Time Span for Concurrent Diagnosis}\label{sec3.4}
\begin{figure}[H]
\centering
\includegraphics[clip,width=0.85\columnwidth]{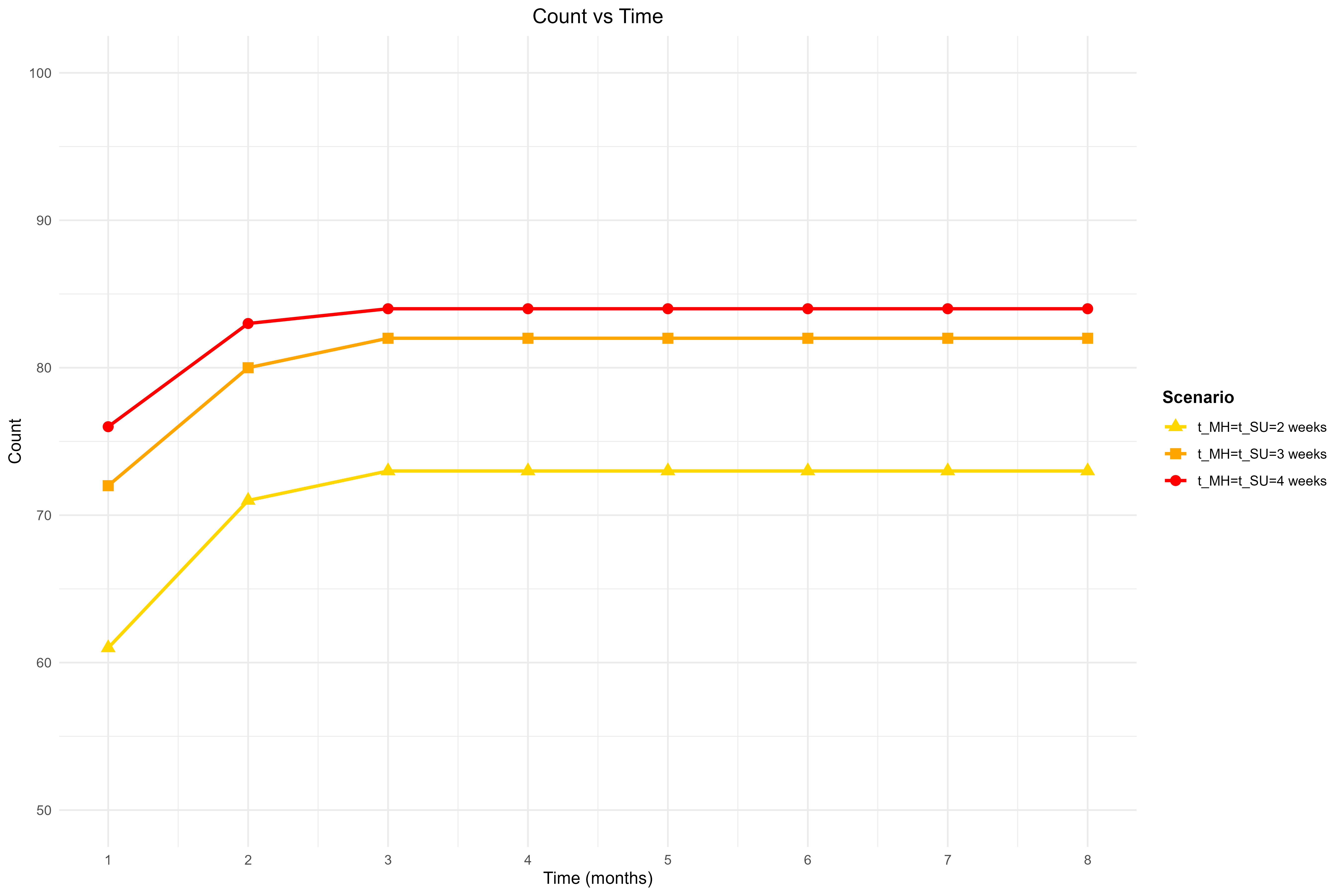}
\caption{Frequency of concurrent MHSU diagnosis vs.\ maximum time span between MH and SU (Time=\texttt{t\_MHSU}), for \texttt{t\_MH} = \texttt{t\_SU} = \{14,21,28\} days. \label{fig6}}
\end{figure}

Finally, we study how varying \texttt{t\_MHSU} affects the frequency of concurrent MHSU detection, assuming fixed maximum spans \texttt{t\_MH} and \texttt{t\_SU}. We require at least two hospital or two physician visits within $x$ days ($x \in \{14,21,28\}$) for MH and SU:
\[
\texttt{n\_MHH} = \texttt{n\_MHP} = \texttt{n\_SUH} = \texttt{n\_SUP}=2,\quad
\texttt{t\_MH} = \texttt{t\_SU} = x,\ \ x=14,21,28,
\]
and then let $\texttt{t\_MHSU}=y$ vary in discrete steps of $31k$ ($k=1,\ldots,12$). Figure~\ref{fig6} shows the results. For each fixed maximum time span for MH and SU, the number of detected patients increases with increasing maximum time span for MHSU diagnosis, approaching an asymptote. Also, for a fixed maximum time span for MHSU diagnosis, increasing  maximum time span for MH and SU leads to capturing more patients.

\subsection{Temporal Analysis}\label{sec3.5}

In this section, we assume that, unlike the previous three sections (\ref{sec3.2}-\ref{sec3.4}), the study's principal investigators have established agreed-upon default values for key parameters. These include a minimum of one hospital visit, a minimum of two medical service physician visits, a maximum time span of one month for mental health diagnoses, a maximum time span of one month for substance use diagnoses, and a maximum time span of one month for concurrent diagnoses. Here, policymakers are particularly interested in monitoring the monthly trends in the frequency and rate of MH, SU, and MHSU diagnoses throughout the 2024 calendar year. Figure~\ref{fig7} presents the results using similar programming presented in Appendix (Sections ``A1. Scalability with large databases" and ``A2. Summary Statistics Output"). As shown in the figure, both MH and SU exhibit an overall increasing trend over time, reaching their peak in November. In contrast, MHSU follows a fluctuating pattern initially, followed by a period of stabilization until November. However, all three categories experience a sharp decline in December.\par 
\begin{rema}
The temporal analysis presented here is based on frequency statistics. A similar approach applies to rate statistics, depending on the objectives of the principal investigators. The key difference lies in using proportions instead of counts in the estimation process, as outlined in Appendix (Section "A2. Summary Statistics Output").
\end{rema}

\begin{rema} Given the nature of the healthcare administrative database and the study objectives, principal investigators have the flexibility to consider various temporal components for [Unit, Span] in the temporal analysis. In this study, our temporal analysis was conducted to examine monthly variations over a year, represented as [Month, Year]. Other potential examples include  [Day, Month], [Week, Year], [Quarter, Year], and [Year, Decade]. 
\end{rema}

\begin{figure}[H]
\centering
\includegraphics[clip,width=0.85\columnwidth]{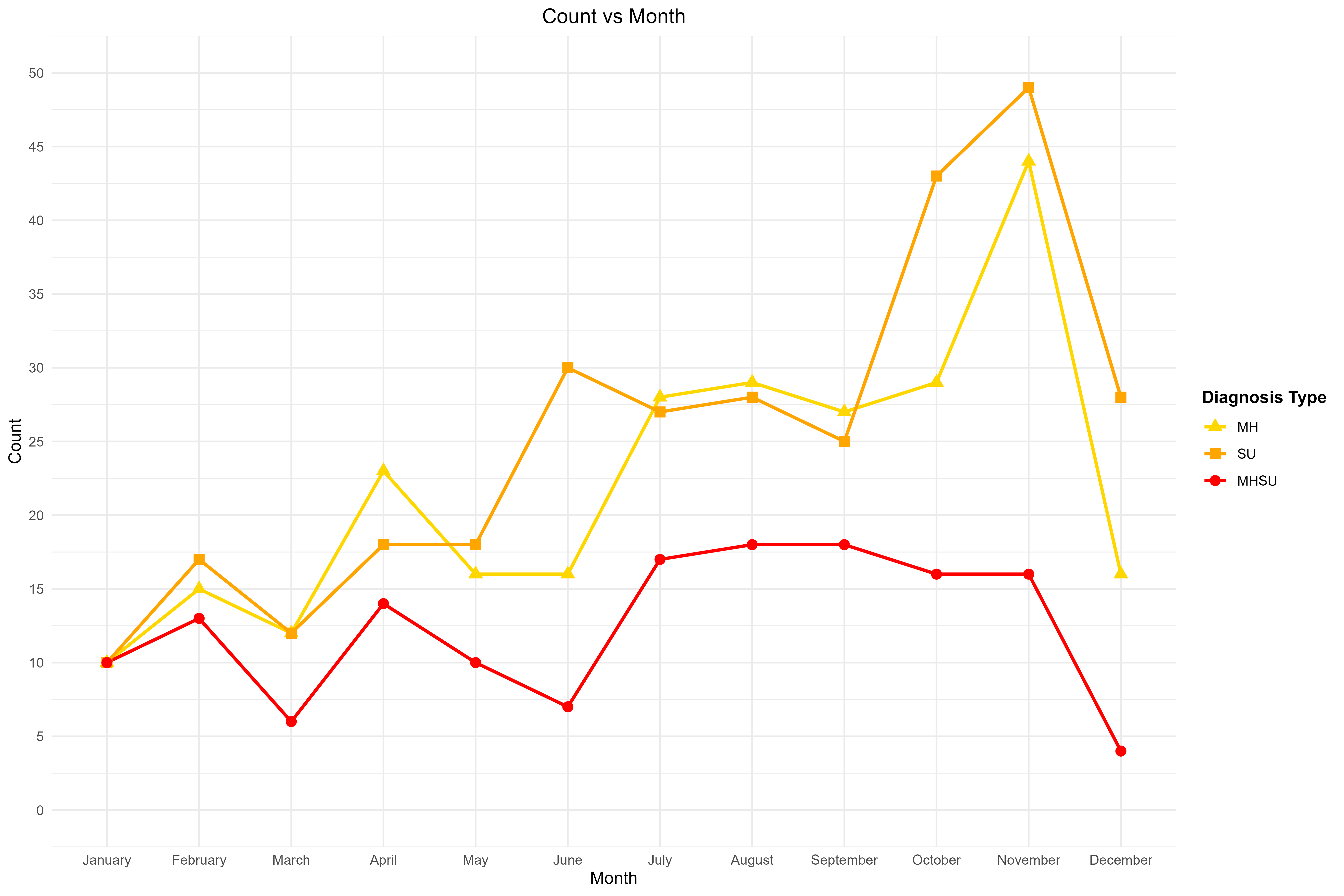}
\caption{Monthly Frequency of MH, SU, and MHSU status in the calendar year 2024. \label{fig7}}
\end{figure}

\section{Discussion}\label{sec4}

\subsection{Summary and Contributions}\label{sec4.1}

We introduce the R package CMHSU, the first statistical software package to implement the Data-Driven Diagnostic Method (DDDM) for identifying mental health (MH), substance use (SU), and concurrent (MHSU) statuses within North American healthcare administrative databases. The package provides flexibility to accommodate various scenarios, including the minimum required visits to hospitals or medical service physicians and the maximum time spans for MH and SU diagnoses and between them.\par 

The first key contribution of the CMHSU package is enabling clinicians to define concurrent MHSU status using the DDDM approach. The examples presented in this paper highlight three critical dimensions and their associated challenges in achieving this definition. Once these dimensions are addressed simultaneously, defining concurrent MHSU status based on DDDM becomes feasible. The details of these dimensions are as follows:\par 
\begin{enumerate}
\item Dimension of Time Span within MH and SU Diagnoses (Section \ref{sec3.2}):
The time span within MH and SU diagnoses plays a critical role. For in-stance, if a maximum time span of 7 days is considered, only 48 out of 100 patients (48.8\%) are captured, whereas extending the time span to 56 days captures 90 out of 100 patients (90.0\%). This delicate situation raises the question: "What is the appropriate maximum time span for MH and SU diagnoses?"\par 
\item Dimension of Required Number of Visits (Section \ref{sec3.3}):
The number of required hospital and medical service physician visits significantly influences the detection rates. For example, requiring two hospital visits and four physician visits captures 90 out of 100 patients (90.0\%), while increasing the requirement to three hospital visits and six physician visits reduces the capture rate to 70 out of 100 patients (70.0\%). This delicate balance prompts the question: "What is the optimal number (or ratio) of visits  needed?"\par 
\item Dimension of Time Span for Concurrent Status (Section \ref{sec3.4}):
The maximum time span between MH and SU diagnoses also plays a vital role. For instance, setting this span to one month captures 61 out of 100 patients (61.0\%), while extending it to three months captures 84 out of 100 patients (84.0\%). This scenario raises the question: "What is the ideal maximum time span between MH and SU diagnoses?"\par 

\end{enumerate}
The second key contribution of the CMHSU package is its ability to diagnose comorbidities of various diseases within healthcare administrative databases. While originally designed for concurrent mental health and substance use diagnoses, the package does not impose limitations on its input ICD variables, making it applicable for other comorbidity conditions.\par 

The final key contribution of the CMHSU package is its utility for policy-makers at federal, state, and territorial levels. By facilitating the rapid compilation of databases tracking ongoing mental health, substance use, and concurrent statuses, the package enables timely monitoring of trends. This real-time capability supports the development and implementation of appropriate healthcare policies in a timely manner.\par 

\subsection{Strengths, Limitations and Future Work }\label{sec4.2}
The R package CMHSU offers several unique features that make it a valuable and advantageous statistical tool for detecting mental health (MH), substance use (SU), and concurrent mental health and substance use (MHSU) status within large administrative healthcare databases. These features include:\par 

\begin{enumerate}
\item Flexibility: CMHSU incorporates four core functions and ten customizable parameters, allowing researchers to account for a wide range of predefined scenarios by investigators when identifying MH, SU, and MHSU status in healthcare administrative databases.\par
\item Comprehensiveness:  CMHSU enables detection of almost all mental health and substance use conditions than other statistical tools given their recent comprehensive codifications in ICD-10 and ICD-11.\par 
\item Efficiency: The CMHSU package is designed for ease of use, requiring only a basic statistical background. Unlike advanced statistical and machine learning-based detection methods for MH\cite{Liuetal2023}, SU\cite{de Mattosetal2024}, and MHSU\cite{Acharyaetal2024}— which typically necessitate expertise in topics such as K-Nearest Neighbors, Random Forests, Gradient Boosted Trees, Deep Neural Networks— CMHSU offers a user-friendly and time-efficient alternative, making it particularly accessible to researchers and practitioners without extensive statistical or machine learning expertise.\par 
\item Interpretability: CMHSU employs a trace-back methodology to identify MH, SU, and MHSU status based on their corresponding ICD codes. This approach enhances the clarity and transparency of the results, facilitating faster interpretation, improved visualization of detected cases, and analysis of trends over time compared to other statistical tools.\par 
\item Seamless Integration: The package is free and easy to install, ensuring compatibility with existing analytical tools used for processing large healthcare administrative databases. This seamless integration enhances its accessibility and usability in real-world research applications.\par 
\end{enumerate}

This work has several limitations, each of which presents opportunities for future extensions. Some of them are as follows:
\begin{enumerate}

\item  Scalability: For large-scale databases, it may be more efficient to partition the input dataset into multiple disjoint subsets and apply the MHSU\_status\_broad() function to each subset separately to manage the extensive outputs. Introducing an additional parameter within the function to automate the partitioning of the input data ($inputdata$) would further optimize the computational process(See Appendix “A1. Scalability with large databases” ).\par 

\item Output Format: The package currently generates outputs as dataframes only, without providing summary frequency statistics for mental health, substance use, and their concurrent statuses. Calculating these statistics requires additional R programming (See Appendix “A2. Summary Statistics Output”).\par 

\item Customization: The window lag in the function MHSU\_status\_broad() is fixed at one day (as illustrated in Figure \ref{fig2}). Adding a fixed or adaptive parameter to specify the length of this window lag would enhance the flexibility of the detection process for researchers. The choice of fixed or adaptive status depends to specific patient data in the study.\par 

\item Evaluation: The DDDM approach assumes that ICD coding in administrative databases is reliable. However, in practice, these records are often incomplete or subject to misclassification, leading to potential biases in the evaluation process and, consequently, affecting the accuracy of the results. Furthermore, it is still unclear how to measure the package's precision in detecting patients with MH status, SU status, and MHSU status, as well as how to compare its performance using DDDM with more advanced machine learning-based methods \cite{Liuetal2023,de Mattosetal2024,Acharyaetal2024}. This comparison needs a mapping mechanism between the above parameters in the DDDM and above machine learning-based methods requiring subsequent methodological research.\par 

\item Standardization: The use of ICD codes for diagnosis depends on local and national jurisdictions (Table \ref{tab1}). Potential inconsistencies in coding across different jurisdictions and healthcare databases may affect the reliability of the results for comparisons across these jurisdictions.\par 

\item Empirical Validation: The simulation study presented in this paper utilizes self-generated simulated data. Ideally, using a real-life empirical healthcare administrative database would significantly enhance the validity and applicability of the findings for healthcare policy implementations. However, key obstacles—including (1) data privacy and security regulations, (2) access restrictions and bureaucratic hurdles, and (3) selection bias—prevented the use of such database in the current analysis.\par 

\item Geographical Adaptability: CMHSU is designed for detecting mental health (MH) status , substance use (SU) status , and their concurrent (MHSU) status  within the North American healthcare databases, based on the fact that the original DDDM was developed by researchers in this region. However, it remains unclear whether these methods can be effectively adopted for use in the non-North American healthcare systems, given potential differences in healthcare infrastructure, coding practices, and administrative data structures. Despite all these issues, given availability of the essential data fields, a preliminary analysis is still possible (See Section \ref{sec4.3}).\par

\item Temporal Analysis: The simulation study in this paper serves as a hypothetical demonstration of the package's application. Its results can only be truly meaningful and applicable to real-world policymaking if the DDDM parameters were based on a universally agreed-upon set of predefined values related to time span and visit definitions. However, no such universally accepted standard parameters currently exists among psychiatrists. Addressing this issue requires further discussions, research and a final systematic review and meta-analysis among various studies. \par 
\end{enumerate}

\subsection{Summary of CMHSU Data Analysis Workflow}\label{sec4.3}

We conclude this section with a summary of the data analysis workflow, designed to facilitate a smooth user experience when applying the CMHSU package in their projects. The workflow comprises the following steps:

\begin{enumerate}[label={(\roman*)}, leftmargin=*, itemindent=1em]
\item  Compile CMHSU Data:\\
Gather essential data fields, including $ClientID, VisitDate, Diagnostic_H, and,\  Diagnostic_P.$
\item  Assess Scalability:
\begin{itemize}
    \item[(a)] \textit{Size}: Execute data size scalability by applying the function \texttt{splitfunction\_id()}.
    \item[(b)] \textit{Time}: Execute temporal scalability by applying the function \texttt{splitfunction\_time()}.
\end{itemize}
\item  Define Analysis Parameters:\\
Specify required parameters, such as \texttt{n\_MHH}, \texttt{n\_MHP}, and others as necessary.

\item  Examine Data Time Span:
\begin{itemize}
    \item[(a)] For data spans less than or equal to \texttt{t\_MHSU}, apply the function \texttt{MHSU\_status\_basic()}.
    \item[(b)] For data spans exceeding \texttt{t\_MHSU}, apply the function \texttt{MHSU\_status\_broad()}.
\end{itemize}

\item  Report Results:
\begin{itemize}
    \item[(a)] \textit{Frequency}: Extract count data using the script \texttt{SummarySampleMHSU\_1}.
    \item[(b)] \textit{Proportion}: Extract proportion data using the script \texttt{SummarySampleMHSU\_1}.
\end{itemize}

\item  Temporal Interpretation:\\
Interpret temporal patterns by applying the [Unit, Span] methodology illustrated in Section \ref{sec3.5}.
\end{enumerate}
 
\section{Conclusions}\label{sec5}

The free R software package CMHSU enables researchers to detect mental health (MH) status, substance use (SU) status, and their concurrent (MHSU) status within healthcare administrative databases. The package offers a wide range of flexibility regarding visit count and maximum time span parameters, allowing for comprehensive and adaptable analyses. This functionality facilitates the compilation of these statuses at a large population level in a timely manner, supporting health policymakers in effectively monitoring trends and making informed healthcare decisions to optimize patient treatment.\par

\vspace{1em}
\noindent
\textbf{Supplementary Material.} 
\subsubsection*{S1. R Package Materials}
The following supporting information can be downloaded at “CMHSU” R package CRAN documentation: \url{https://cran.r-project.org/package=CMHSU}.

\subsubsection*{S2. Shiny Application} 
A Shiny application has been included in the supplementary materials to facilitate interactive exploration of the estimations. The application can be accessed at:
\url{https://mohsensoltanifar.shinyapps.io/rpackageshinyappcmhsu/}.

\vspace{1em}
\noindent
\textbf{Author Contributions.} 
Conceptualization, M.S.; methodology, M.S.; software, M.S., C.H.L.; validation, M.S., C.H.L.; formal analysis, M.S.; investigation, M.S.; resources, M.S., C.H.L.; data curation, M.S.; writing---original draft, M.S., C.H.L.; writing---review and editing, M.S., C.H.L.; visualization, M.S.; supervision, M.S., C.H.L.; project administration, M.S.; funding acquisition, M.S. All authors have read and agreed to the published version of the manuscript.

\vspace{1em}
\noindent
\textbf{Funding.} 
This research received no external funding. The APC of this work has been funded by the first author.

\vspace{1em}
\noindent
\textbf{Acknowledgments.} 
This paper is dedicated to the families and loved ones of those affected by the fentanyl and other illicit drugs  crisis in Canada and the United States. May this work serve as another steppingstone towards meaningful solutions that prevent further loss and offer healing to communities in both nations (\raisebox{-0.25\height}{\includegraphics[height=12pt]{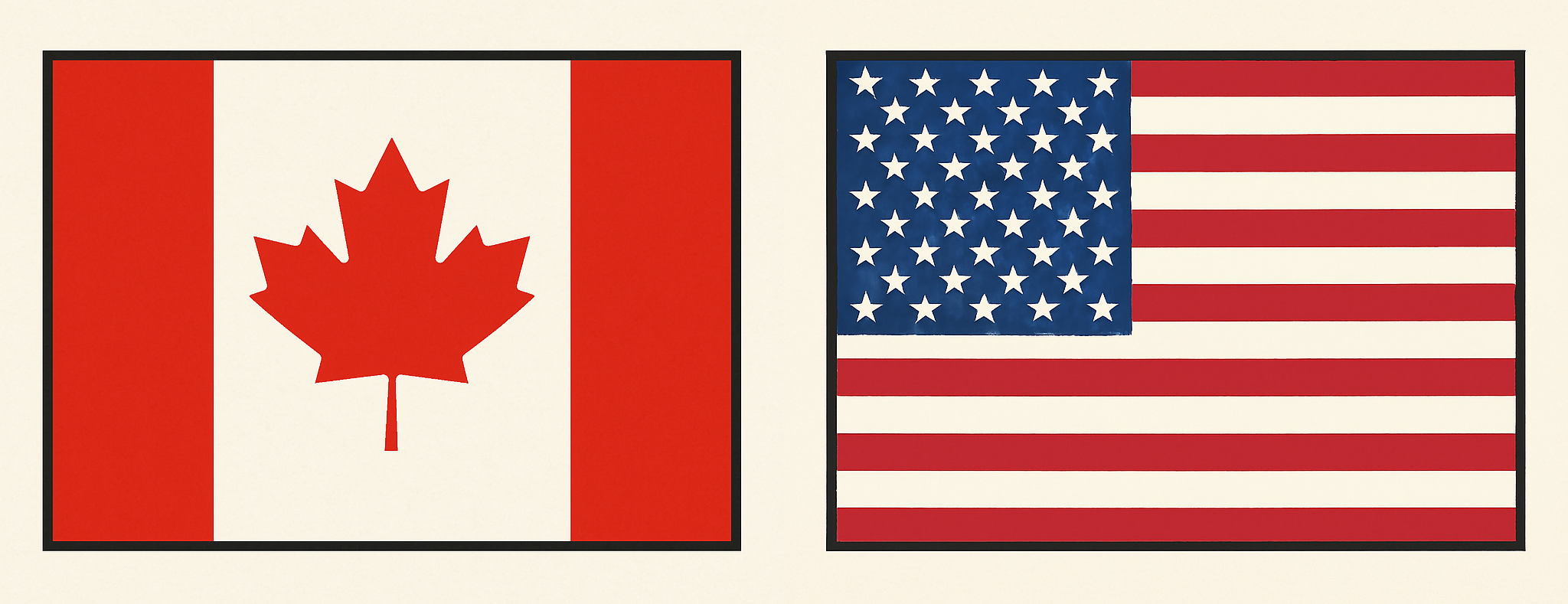}}).\par

\vspace{1em}
\noindent
\textbf{Conflicts of Interest.} 
The authors declare no conflicts of interest.

\vspace{1em}
\noindent
\textbf{Abbreviations.}\\
\noindent
The following abbreviations are used in this manuscript:\\
APCD: All-Payer Claims Databases; AUDIT: Alcohol Use Disorders Identification Test; CBT: Cognitive-Behavioral Therapy; CCA: Comprehensive Clinical Assessment; CMDB: Canadian MIS Database; CRAN: Comprehensive R Archive Network; DAD: Discharge Abstract Database; DAST: Drug Abuse Screening Test; DDDM: Data-Driven Diagnostic Method; GAD-7: Generalized Anxiety Disorder-7; ICD: International Classification of Diseases; ITA: Integrated Treatment Approach; MAT: Medication-Assisted Treatment; MC: Multidisciplinary Collaboration; MEDPAR: Medicare Provider Analysis and Review; MH: Mental Health; MHSU: Concurrent Mental Health and Substance Use; MSP: Medical Services Plan; NACRS: National Ambulatory Care Reporting System; NAMCS: National Ambulatory Medical Care Survey; NIS: National Inpatient Sample; OHIP: Ontario Health Insurance Plan; OSHPD: California Office of Statewide Health Planning and Development; PHQ-9: Patient Health Question-naire-9; RAMQ: Régie de l'assurance maladie du Québec; SID: State Inpatient Databases; SPARCS: New York Statewide Planning and Research Cooperative System; SSI: Standardized Screening Instruments; SU: Substance Use; VA: Veterans Affairs.  
 
\section*{Appendix}
\subsection*{A1. Scalability with large databases :}\label{secA1}
The scalability of the large healthcare administrative dataset, inputdata, is achieved by considering two dimensions: unique IDs and time. First, for the unique ID dimension, the function splitfunction\_id(), introduced below, takes the dataset inputdata (containing $m$ unique patients) and an integer $n\geq1$ as input. It then partitions the dataset into $k$ subsets, where  $k=[m/n]+ 1 (if\ \  n \nmid m), m/n (if\ \  n\mid m).$ The first $k-1$ subsets each contain $n$ unique patients, while the kth subset contains at most $n$ unique patients. For example, we consider the case where $m=200, n=18$ and $k=12:$\par 

\begin{Verbatim} 
line #1:> splitfunction_id <- function(inputdata, n) {
                              inputdata <- inputdata %>% arrange(ClientID)
                              unique_ids <- unique(inputdata$ClientID)
                              groups <- split(unique_ids, ceiling(seq_along(unique_ids)/n))
                              split_datasets <- list()
                              for (i in seq_along(groups)) {
                              split_datasets[[paste0("inputdata_", i)]] <- inputdata %>% 
                              filter(ClientID %in% groups[[i]])
                              }
                              return(split_datasets)   
                              }
line #2:> inputdata_split_id <- splitfunction_id(SampleRWD, 18)
line #3:> inputdata_split_id$inputdata_1
\end{Verbatim}

Second, for the time dimension, the function splitfunction\_time(), introduced below, takes the dataset inputdata (with time span $T$ days) and an integer $t\geq1$ as input. It then partitions the dataset into $l$ subsets, where  $l=[T/t]+ 1 (if\ \  t \nmid T), T/t (if\ \  t\mid T).$ The first $l-1$ subsets each are in the time span of $t$ days, while the lth subset  time span is at most $t$. For example, we consider the case where $T=363, t=30.5$ and $l=12:$\par 

\begin{Verbatim}
line #1:> splitfunction_time <- function(inputdata, t) {
                                inputdata <- inputdata %>% mutate(VisitDate = as.Date(VisitDate))
                                inputdata <- inputdata %>% arrange(VisitDate)
                                VisitDate_0 <- min(inputdata$VisitDate, na.rm = TRUE)
                                split_datasets <- list()
                                i <- 1
                                while(TRUE) {
                                start_date <- VisitDate_0 + (i - 1) * (t + 1)
                                end_date <- start_date + t
                                subset_data <- inputdata %>% 
                                filter(VisitDate >= start_date & VisitDate <= end_date)
                                if (nrow(subset_data) == 0) break  
                                split_datasets[[paste0("inputdata_", i)]] <- subset_data
                                i <- i + 1
                                }
                                return(split_datasets)   
                                }
line #2:> inputdata_split_time <- splitfunction_time(SampleRWD, 30.5)
line #3:> inputdata_split_time$inputdata_1
\end{Verbatim}

\subsection*{A2. Summary Statistics Output:}\label{secA2}
The following R program computes summary statistics for MH, SU, and MHSU frequencies from the CMHSU output shown in the Figure \ref{fig3} (c):
\begin{Verbatim} 
line #1:> myexample<-SampleRWD[,c(1:4)]
line #2:> SampleMHSU_1 = MHSU_status_basic(myexample, n_MHH=1, n_MHP=1, n_SUH=1,
                 n_SUP=1, t_MH=60, t_SU=60, t_MHSU=365,
                 ICD_MH=c("F060","F063","F064","F067"),
                 ICD_SU=c("F100","T4041","F120","F140"))
line #3:> SummarySampleMHSU_1 <-  SampleMHSU_1 %>%
          summarise(
            MH_Count = sum(MH_status == "YES"),
            MH_Proportion = formatC(mean(MH_status == "YES"), format = "f", digits = 3),
            SU_Count = sum(SU_status == "YES"),
            SU_Proportion = formatC(mean(SU_status == "YES"), format = "f", digits = 3),
            MHSU_Count = sum(MHSU_status == "YES"),
            MHSU_Proportion = formatC(mean(MHSU_status == "YES"), format = "f", digits = 3) )
line #4:> print(SummarySampleMHSU_1)

line #5:> # A tibble: 1 × 6
          MH_Count MH_Proportion SU_Count SU_Proportion MHSU_Count MHSU_Proportion
          <int>     <chr>        <int>     <chr>       <int>        <chr>
          125       0.625         125      0.625        100         0.500
\end{Verbatim}


\end{document}